# Muon spin rotation and relaxation in Pr$_{1-x}$Nd$_x$Os$_4$Sb$_{12}$: superconductivity and magnetism in Pr-rich alloys

P.-C. Ho,[1] D. E. MacLaughlin,[2] M. B. Maple,[3] Lei Shu,[4] A. D. Hillier,[5] O. O. Bernal,[6] T. Yanagisawa,[7] P. K. Biswas,[5] Jian Zhang,[4] Cheng Tan,[4] S. D. Hishida,[1] and T. McCullough-Hunter[1]

[1]*Department of Physics, California State University, Fresno, California 93740, USA*
[2]*Department of Physics and Astronomy, University of California, Riverside, California 92521, USA*
[3]*Department of Physics, University of California, San Diego, La Jolla, California 92093, USA*
[4]*State Key Laboratory of Surface Physics, Department of Physics, Fudan University, Shanghai 200433, China*
[5]*ISIS Facility, Rutherford Appleton Laboratory, Chilton, Didcot Oxon, OX11 0QX, United Kingdom*
[6]*Department of Physics and Astronomy, California State University, Los Angeles, California 90032, USA*
[7]*Department of Physics, Hokkaido University, Sapporo, Hokkaido 060-0810, Japan*

(Dated: September 18, 2019)

## Abstract

The Pr-rich end of the alloy series $Pr_{1-x}Nd_xOs_4Sb_{12}$ has been studied using muon spin rotation and relaxation. The end compound $PrOs_4Sb_{12}$ is an unconventional heavy-fermion superconductor, which exhibits a spontaneous magnetic field in the superconducting phase associated with broken time-reversal symmetry. No spontaneous field is observed in the Nd-doped alloys for $x \geqslant 0.05$. The superfluid density is insensitive to Nd concentration, and no $Nd^{3+}$ static magnetism is found down to the lowest temperatures of measurement. Together with the slow suppression of the superconducting transition temperature with Nd doping, these results suggest anomalously weak coupling between Nd spins and conduction-band states.

## I. INTRODUCTION

Since the discovery of unconventional superconductivity (SC) in the filled skutterudite compound $PrOs_4Sb_{12}$ ($T_c = 1.85$ K) [1, 2], there are still many intriguing aspects of this compound that are not well understood. Two of these are the origin its enhanced electron mass in the absence of the magnetic crystalline-electric-field (CEF) ground state [3, 4] that is usual for heavy-fermion compounds, and the mechanism or mechanisms responsible for its multiple SC phases [5–7]. An internal magnetic field was detected in the zero-field SC state [8]. This is indicative of broken time reversal symmetry (TRS), which makes it also a candidate for a $p$-wave superconductor. The appearance of SC in $PrOs_4Sb_{12}$ is also close to an antiferroquadrupolar(AFQ) phase for magnetic field $H$ between 4.5 T and 14.5 T [9–11], raising the possibility that the pairing mechanism comes from quantum fluctuations of quadruple moments.

Other rare-earth filled-skutterudite compounds also have ordered states at quite low temperatures: spin-density-wave (SDW) type antiferromagnetism in $CeOs_4Sb_{12}$ occurs below 1 K [12–14], ferromagnetism (FM) in $NdOs_4Sb_{12}$ appears below 1 K [15], and weak FM in $SmOs_4Sb_{12}$ develops below 2.6 K [16]. These effects led to speculation that the origin of the SC pairing mechanism in $PrOs_4Sb_{12}$ could be quantum fluctuations near a magnetic quantum critical point (QCP).

In order to investigate the influence of FM on the unconventional SC in $PrOs_4Sb_{12}$, we have have carried out a study of the $Pr_{1-x}Nd_xOs_4Sb_{12}$ substitutional alloy system [17–19]. Previously we used the muon spin rotation/relaxation ($\mu$SR) technique to study the progress of the FM state for samples with higher concentrations of $x$ [18, 19]. In this report, we focus on our $\mu$SR investigation of the evolution of broken TRS and other properties in Pr-rich $\text{PrOs}_4\text{Sb}_{12}$.

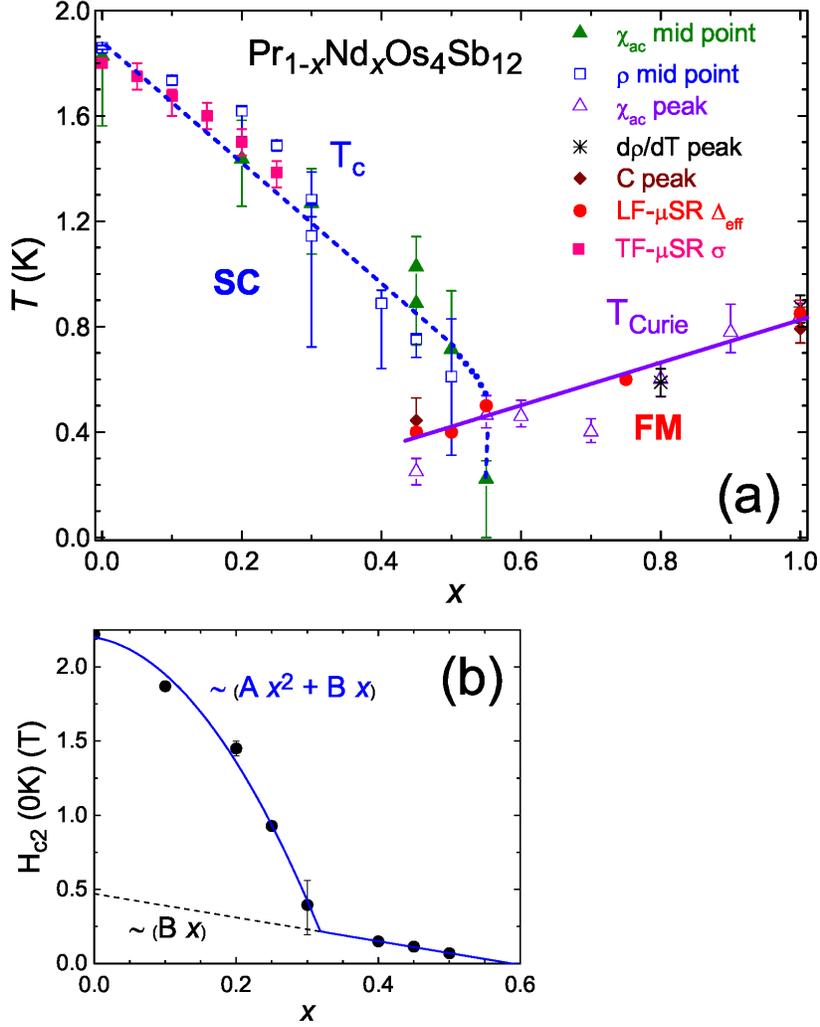

FIG. 1. (a) $T$–$x$ Phase diagram of superconducting transition temperature $T_c$ and ferromagnetic transition temperature $T_{\text{Curie}}$ vs Nd concentration $x$. $\mu$SR data from this paper ($x = 0.05, 0.10, 0.15,$ and $0.20$), Ref. 20 ($x = 0$), and Ref. 18 ($x = 0.25$). Transition temperatures from electrical resistivity $\rho$, magnetic susceptibility $\chi$, and specific heat $C$ [17] and $\mu$SR (this work). (b) $T = 0$ upper critical field $H_{c2}$ vs $x$ for $\text{Pr}_{1-x}\text{Nd}_x\text{Os}_4\text{Sb}_{12}$.

The phase diagram of superconducting transition temperature $T_c$ and ferromagnetic transition temperature $T_{\text{Curie}}$ vs Nd concentration $x$ is shown in Fig. 1(a). Figure 1(b) shows the Nd concentration dependence of the upper critical field $H_{c2}(T=0, x)$ for $Pr_{1-x}Nd_xOs_4Sb_{12}$. $H_{c2}(x)$ is nonzero below a critical concentration $x_{cr1} \approx 0.58$. An anomaly in $H_{c2}$ appears at a second critical concentration $x_{cr2} \approx 0.33$. Weak FM was detected for $x_{cr2} \leq x \leq x_{cr1}$ within the SC state [17].

This article reports results of a study of Pr-rich $Pr_{1-x}Nd_xOs_4Sb_{12}$ alloys, $x = 0, 0.05, 0.10, 0.15, 0.20$, and 0.25, using zero-field and transverse-field $\mu$SR (ZF-$\mu$SR and TF-$\mu$SR, respectively) [21–25]. We find that the spontaneous internal field previously found in the end compound $PrOs_4Sb_{12}$ [8, 20] is rapidly suppressed by Nd doping (critical concentration < 0.05), a surprising result since the $Nd^{3+}$ ions possess $4f$ magnetic moments. At higher Nd concentrations the dynamic relaxation rate increases with decreasing temperature, reminiscent of critical slowing at a magnetic phase transition. The superfluid density, measured using TF-$\mu$SR, is remarkably independent of $x$ up to $\sim$0.2 and then decreases. No magnetic transition is observed down to lowest temperatures of measurement for $x \leq 0.25$, consistent with earlier results.

## II. EXPERIMENT

Single crystals of $Pr_{1-x}Nd_xOs_4Sb_{12}$, $x = 0.05, 0.10, 0.15$, and 0.20, were prepared by the antimony self-flux growth method [12]. Praseodymium and neodymium were premixed using a single-arc furnace. X-ray powder diffraction measurements were made to confirm the samples have the cubic $LaFe_4P_{12}$-type structure [26]. For each concentration $\sim$2-g clusters of single crystals were attached to silver sample holders using GE-7031 varnish.

$\mu$SR experiments were carried out at the ISIS muon facility at the STFC Rutherford Appleton Laboratory over the temperature range 0.025–3 K, in zero magnetic field (ZF) and applied transverse magnetic fields (TF, i.e., perpendicular to the initial muon spin polarization) between 0 and 200 Oe. For the ZF-$\mu$SR experiments the field was zeroed to within $\sim$1 $\mu$T. Previously-reported data for $x = 0$ [20, 27–29] and 0.25 [18], obtained at ISIS and TRIUMF, Vancouver, Canada, have been reanalyzed, and the results are included in the present report.

In a typical $\mu$SR experiment [21–25], 100% spin-polarized positive muons ($\mu^+$) are implanted in the samples and come to rest at interstitial sites. Each muon decays according to the reaction $\mu^+ \to e^+ + \nu_e + \overline{\nu}_\mu$, with a lifetime $\tau_\mu \approx 2.197$ $\mu$s. Decay positrons are emitted preferentially in the direction of the $\mu^+$ spin at the time of decay, and are detected using an array of scintillation counters surrounding the sample. In a TF-$\mu$SR experiment counters are aligned perpendicular to the field. In ZF-$\mu$SR counters are aligned parallel to the initial polarization; two counters surrounding the sample ["forward" ($f$) and "back" ($b$)] are generally used.

For each counter the positron count rate is given by

$$N(t) = N^0 e^{-t/\tau_\mu} \left[1 + AG(t)\right], \quad (1)$$

where $G(t)$ is the time evolution of the $\mu^+$ ensemble spin polarization component in the direction of the counter ($G = 1$ for 100% polarization), $N^0$ is the initial count rate, and $A$ is the count-rate asymmetry, spectrometer-dependent but usually $\sim$0.2–0.25 [21–25]. For two identical counters $f$ and $b$, oriented 180° ($N_b^0 = N_f^0$, $A_b = -A_f$), the asymmetry spectrum

$$A(t) = \frac{N_f(t) - N_b(t)}{N_f(t) + N_b(t)} = AG(t) \quad (2)$$

contains the essential information conveyed by muons concerning local static and dynamic magnetic properties of the sample. In practice the two counters and their environments are seldom identical, but differences can be measured in calibration experiments and accounted for in data analysis [30]. In ZF-$\mu$SR $G(t=0) = 1$.

A spurious signal is often present due to muons that miss the sample and stop in the sample holder. This signal has been determined from the fits and is subtracted to yield the signal from the sample.

### A. Zero Field

In ZF the relaxation function $G(t)$ is extremely sensitive to the behavior of the local magnetic field $\mathbf{B}_{\rm loc}$ at $\mu^+$ sites, since there is no competing applied field. If dynamic relaxation is negligible, $\mu^+$ precession in $\mathbf{B}_{\rm loc}$ results in static relaxation of $G(t)$ if the magnitude $B_{\rm loc}$ is distributed. The spin component of each muon parallel to $\mathbf{B}_{\rm loc}$ does not precess, however, leading to a constant term in $G(t)$ [31]. If $\mathbf{B}_{\rm loc}$ is randomly oriented, in zero applied field the component of non-precessing $\mu^+$ polarization in the initial polarization direction is 1/3 the initial polarization [21–25], so that $G(t)$ relaxes from 1 at $t = 0$ to 1/3 at late times.

In the static Gaussian Kubo-Toyabe (KT) model [31] the distribution of static field components is modeled by a Gaussian of width $\Delta/\gamma_\mu$ and zero mean, where $\gamma_\mu = 8.5156 \times$ $10^8$ s$^{-1}$ T$^{-1}$ is the muon gyromagnetic ratio. In ZF the static KT $\mu^+$ spin relaxation function resulting from this field distribution is

$$G_G(\Delta, t) = \frac{1}{3} + \frac{2}{3}\left[1 - (\Delta t)^2\right] \exp\left[-\tfrac{1}{2}(\Delta t)^2\right]. \tag{3}$$

This model is applicable when $\mathbf{B}_{\text{loc}}$ is the sum of randomly-oriented quasistatic nuclear dipolar fields at $\mu^+$ sites if, as is often the case, nuclear moment fluctuations are slow on the muon time scale [31]. The dipole field distribution is approximately Gaussian because random fields from several nuclei contribute and the central limit theorem is approximately valid. The static Gaussian KT model also describes additional (Gaussian-distributed) static contributions to $\mathbf{B}_{\text{loc}}$, such as the spontaneous field observed below $T_c$ in PrOs$_4$Sb$_{12}$ by $\mu$SR [8].

In dilute spin glasses the local field distribution due to impurity spins is Lorentzian in certain limits [32, 33]. The corresponding static KT relaxation function in ZF is [34]

$$G_L(a, t) = \frac{1}{3} + \frac{2}{3}(1 - at)\exp(-at), \tag{4}$$

where $a/\gamma_\mu$ is the half-width of the Lorentzian distribution. More generally, local field distributions in disordered systems where there is more weight in the wings than for a Gaussian may often be approximated, fully or in part, by a Lorentzian. A convolution of Gaussian and Lorentzian distributions yields the ZF static Voigtian KT relaxation function [35]

$$G_V(\Delta, a, t) = \frac{1}{3} + \frac{2}{3}\left[1 - at - (\Delta t)^2\right] \\ \times \exp\left[-at - \tfrac{1}{2}(\Delta t)^2\right]. \tag{5}$$

Dynamic muon spin relaxation (the spin-lattice relaxation of NMR) is due to thermally-fluctuating components of $\mathbf{B}_{\text{loc}}$, which induce transitions between muon spin states and relax $G(t)$ to its thermal equilibrium value (essentially zero in $\mu$SR). A damping factor $\exp(-\lambda_d t)$ is often used to model dynamic relaxation in the presence of static relaxation; thus

$$G(t) = e^{-\lambda_d t}\, G_{\text{stat}}(t), \tag{6}$$

where $G_{\text{stat}}(t)$ is the appropriate static relaxation function. The exponential form of the damping assumes that the fluctuating field at each $\mu^+$ site results in the same dynamic relaxation rate, even though the static local fields are distributed, and that the fluctuation rate is rapid compared to the rms amplitude of the fluctuating field; this is the so-called motionally-narrowed limit. Exponentially-damped Voigtian relaxation [Eq. (6) with $G_{\text{stat}}(t) = G_V(\Delta, a, t)$] has been reported in the filled skutterudite superconductor PrPt$_4$Ge$_{12}$ [36], and is thus a candidate for describing the present experimental results.

### B. Transverse Field

In TF-$\mu$SR $G(t)$ is dominated by $\mu^+$ precession in the applied field, which is damped by relaxation from both static and dynamic mechanisms. For both exponential and Gaussian contributions to the damping, a typical TF relaxation function is

$$G(t) = \exp\left[-\lambda t - \tfrac{1}{2}(\sigma t)^2\right] \cos(\omega t + \phi) \qquad (7)$$

for TF field $H_T \gg B_{\text{loc}}$, where $\omega = \gamma_\mu H_T$, $\sigma/\gamma_\mu$ is the width of the Gaussian field distribution, and $\phi$ is the phase of the oscillation, determined by the geometry of the spectrometer. Whether the exponential TF rate $\lambda$ in Eq. (7) is static or dynamic in origin cannot be determined from TF-$\mu$SR alone.

## III. RESULTS

### A. Zero Field

Observed ZF-$\mu$SR spin polarization functions in Pr$_{1-x}$Nd$_x$Os$_4$Sb$_{12}$ at $T = 1$ K are shown in Fig. 2 for $x = 0.05$, 0.10, and 0.20. The spurious sample-holder signal mentioned in Sec. II has been subtracted, and the data are normalized to 1 at $t = 0$ [cf. Eq (2)] to yield the $\mu^+$ spin relaxation function $G(t)$.

It is not easy to differentiate between static Voigtian relaxation and exponentially-damped static (Gaussian or Voigtian) relaxation if only early-time data are available. If, however, the data extend to late enough times, the difference in asymptotic behavior provides a clear determination. In general $G(t\to\infty) \to 1/3$ for static ZF relaxation due to randomly-oriented local fields (The KT models of Sec. II A are examples), but $G(t\to\infty) \to 0$ for dynamic relaxation. The solid curves in Fig. 2 are fits of exponentially-damped Voigtian KT relaxation functions [Eq. (6) with $G_{\text{stat}}(t) = G_V(\Delta, a, t)$] to the data. The dashed curves are not fits; they are undamped Voigtian KT functions [Eq. (5)] with rates $\Delta$ and $a = \lambda_d$ fixed at values from damped Voigtian fits. They exhibit the expected late-time asymptotic approach to 1/3 for static ZF relaxation and disagree with the experimental data. Furthermore, values of the static Lorentzian rate $a$ from the exponentially-damped Voigtian KT fits are close to zero (Fig. 4). The observed decay to zero is strong evidence that the relaxation is dynamic.

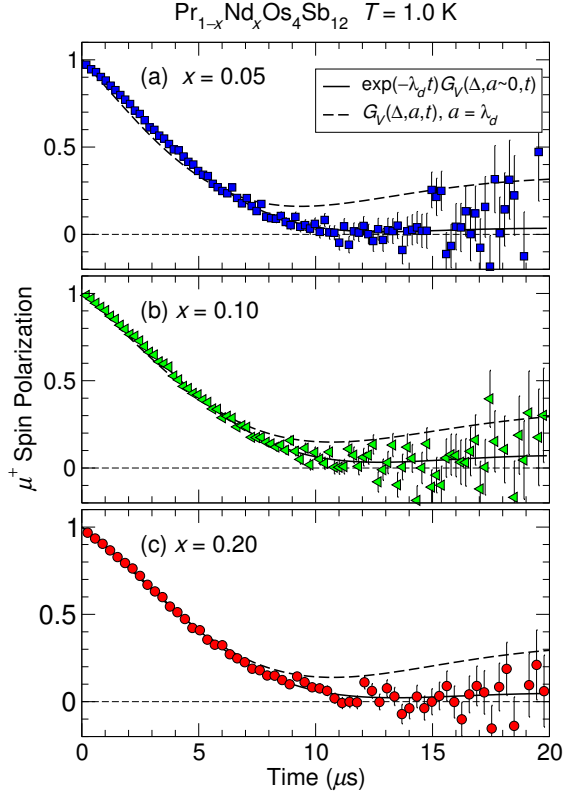

FIG. 2. Zero-field $\mu^+$ spin relaxation in $\mathrm{Pr}_{1-x}\mathrm{Nd}_x\mathrm{Os}_4\mathrm{Sb}_{12}$, $T = 1$ K. (a) $x = 0.05$. (b) $x = 0.10$. (c) $x = 0.20$. Solid curves: fits of damped static ZF Voigtian KT functions [Eq. (6) with $G_{\mathrm{stat}}(t) = G_V(\Delta, a, t)$] to the data (fit value of $a \approx 0$). Dashed curves: plots (not fits) of undamped Voigtian functions [Eq. (5)] with $\Delta$ and $a = \lambda_d$ from damped Voigtian fits.

Figure 3 gives $\mu^+$ spin polarization functions at the lowest temperatures of measurement for representative Nd concentrations. The relaxation initially slows with increasing $x$ as the Gaussian spontaneous field distribution associated with broken TRS is suppressed. For higher Nd concentrations the dynamic relaxation is faster and more nearly exponential.

Figure 4 gives the temperature dependence of the ZF rate parameters for $\mathrm{Pr}_{1-x}\mathrm{Nd}_x\mathrm{Os}_4\mathrm{Sb}_{12}$, $x = 0$, 0.05, 0.10, 0.15, 0.20, and 0.25, from fits of Eq. (6) to the data. In the fits $a$ is a free parameter for $x = 0$, but is fixed at the normal-state average for the Nd-doped alloys. Data have been previously reported for $x = 0$ [20] and $x = 0.25$ [18], where the fits used the static ZF Gaussian function [Eq. (5) with $a = 0$] for $G_{\text{stat}}(t)$. The results of the Voigtian fits in Fig. 4 are essentially the same, since $a \approx 0$ when left free in the fits.

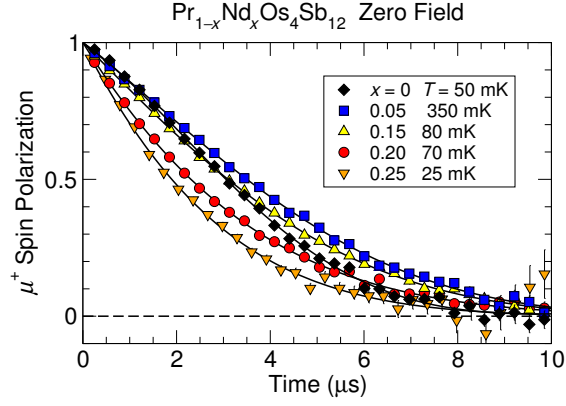

FIG. 3. Zero-field relaxation of $\mu^+$ spin polarization in $Pr_{1-x}Nd_xOs_4Sb_{12}$, $x = 0, 0.5, 0.15, 0.20$ and 0.25, at the lowest temperatures of measurement. Results for $x = 0$ and 0.25 are from reanalysis of previously reported data (Refs. 20 and 18, respectively).

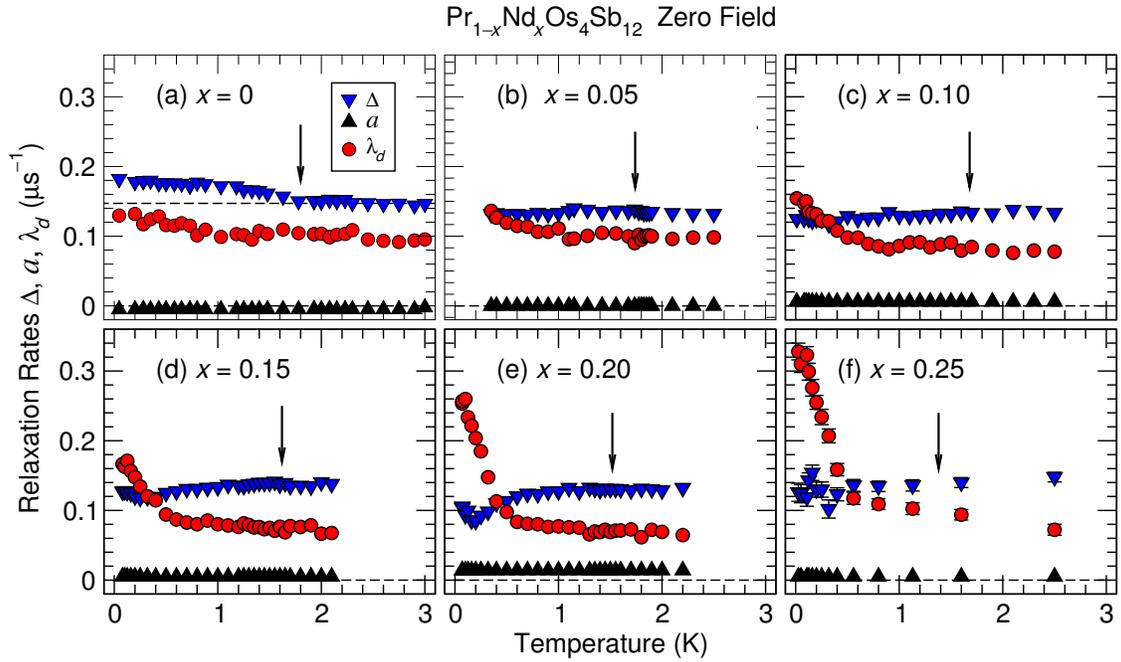

FIG. 4. Temperature dependence of zero-field static Gaussian rate $\Delta$, static exponential rate $a$, and dynamic exponential rate $\lambda_d$ from fits of Eq. (6) to ZF data, $x = 0, 0.5, 0.10, 0.15, 0.20$, and 0.25. Arrows: superconducting transition temperatures $T_c$. Results for (a) $x = 0$ and (f) $x = 0.25$ are from reanalysis of previously-reported data (Refs. 20 and 18, respectively).

For $x = 0$ the Gaussian static rate $\Delta$ [Fig. 4(a)] increases below $T_c$. This is due to the spontaneous local field that indicates broken TRS in the superconducting state [8, 20, 37]. In contrast, Figs. 4(b)–(f) show no increase of $\Delta$ below $T_c$, so that $x = 0.05$ is an upper bound on the critical concentration for suppression of the spontaneous field.

For $x = 0$ the dynamic relaxation rate $\lambda_d$ is significant, and has been attributed to hyperfine-enhanced $^{141}$Pr nuclear spin dynamics [28]. In the alloys $\lambda_d$ exhibits an further upturn at low temperatures, the size of which increases with increasing $x$. This is evidence that Nd$^{3+}$ fluctuations are slowing down, reminiscent of critical slowing associated with a magnetic phase transition. There is, however, no sign of static Nd$^{3+}$ magnetism (e.g., an increase of the static rate $\Delta$) for $x \lesssim 0.25$ down to the lowest temperatures of measurement.

Indeed, for higher Nd concentrations $\Delta(T)$ decreases at low temperatures. This might be expected: slowing of Nd$^{3+}$ fluctuations increases the dynamic relaxation rate of nuclear spins (mainly $^{121}$Sb and $^{123}$Sb), so that nuclear dipolar fields at $\mu^+$ sites are no longer quasistatic. This in turn motionally narrows the $\mu^+$ static relaxation, i.e., reduces the relaxation rate. Presumably the $^{141}$Pr contribution to $\lambda_d(T)$ is also reduced, but increased direct relaxation by Nd$^{3+}$ spins dominates.

### B. Transverse Field

Examples of TF-$\mu$SR spin polarization functions above and below $T_c$ for transverse field $H_T$ greater than the lower critical field $H_{c1}$ are shown in Fig. 5 for $x = 0.05$. The rapid damping for $T = 1.0$ K is due to the inhomogeneous magnetic field distribution in the vortex lattice. This distribution is not expected to be Gaussian [38, 39], but the relaxation rate $\sigma$ from a Gaussian fit is often taken to be given by $\gamma_\mu \langle \delta B^2 \rangle^{1/2}$, where $\langle \delta B^2 \rangle^{1/2}$ is the rms width of the field distribution in the vortex lattice. For $H_T$ close to $H_{c1}$, Brandt [40] showed that

$$\langle \delta B^2 \rangle = 0.00371\, \Phi_0^2/\Lambda^4 \tag{8}$$

in the London limit $\Lambda \gg$ coherence length $\xi$, where $\Lambda$ is the London penetration depth and $\Phi_0$ is the flux quantum. In turn, from the London equations $\Lambda$ is related to the superfluid

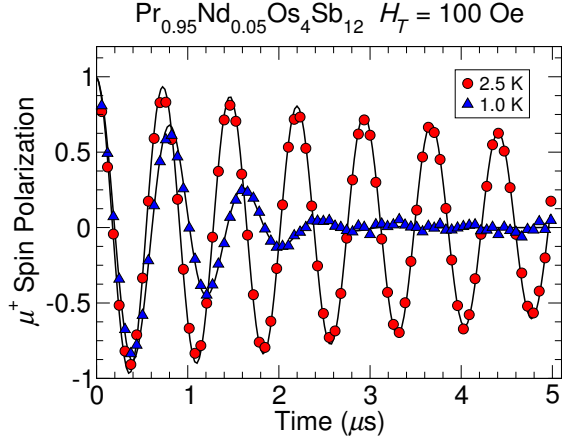

FIG. 5. Time evolution of $\mu^+$ spin polarization for $Pr_{0.95}Nd_{0.05}Os_4Sb_{12}$, transverse field $H_T = 100$ Oe. Circles: $T = 2.5$ K $> T_c$. Triangles: 1.0 K $< T_c$. Curves: fits of Eq. (7) to the data.

density $\rho_s$ and the carrier effective mass $m_{\text{eff}}$ by

$$\frac{1}{\Lambda^2} = \frac{4\pi e^2 \rho_s}{m_{\text{eff}} c^2}. \tag{9}$$

Thus $\langle \delta B^2 \rangle^{1/2} \propto 1/\Lambda^2$ is a measure of the superfluid density.

Figure 6 gives the temperature dependence of the TF rate parameters $\sigma$ and $\lambda$ for $x = 0$, 0.05, 0.10, 0.15, 0.20, and 0.25, from fits of Eq. (7) to the data. The zero-temperature Gaussian rate $\sigma(0)$, and thus the corresponding superfluid density $\rho_s(0)$, varies somewhat with $x$ but is remarkably insensitive to Nd doping up to $x = 0.20$.

For $x = 0$ the TF exponential damping rate $\lambda(T)$ is nearly constant (and approximately the same as the LF rate, cf. Fig. 4), but it acquires considerable structure in the alloys. The two main features are peaks just below $T_c$ (arrows in Fig. 6), and increases with decreasing temperature at low temperatures. Both of these anomalies increase with Nd concentration.

A peak at a magnetic transition is expected from critical slowing down of dynamic spin fluctuations. The data also resemble the "coherence peak" due to conduction-electron spin-lattice (dynamic) relaxation in a fully-gapped superconductor [41]. Peaks are not observed in ZF $\lambda_d(T)$, however (Fig. 4), so dynamic relaxation is unlikely, and there is no evidence for a magnetic transition near $T_c$ in these alloys. We speculate that the peaks are due to a Lorentzian-like component of the static local fields due to inhomogeneity in $T_c$, which disappears when the entire sample volume is superconducting. In contrast, the low-temperature increase of $\lambda(T)$ is also observed in ZF $\lambda_d(T)$ (Fig. 4), consistent with slowing down of dynamic spin fluctuations.

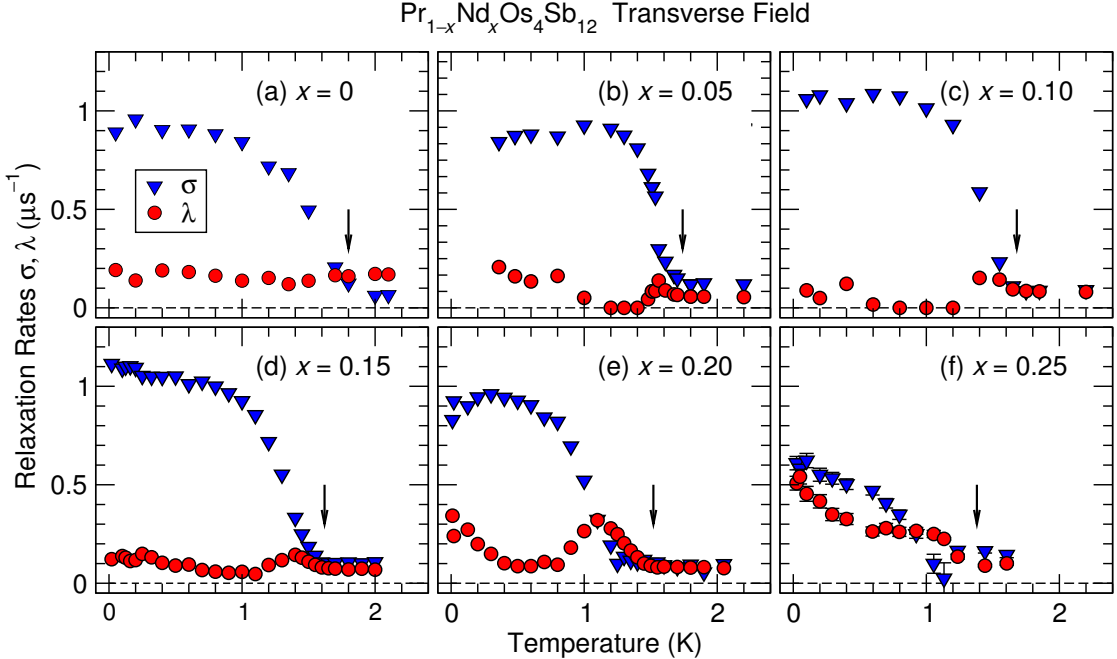

FIG. 6. Temperature dependence of transverse-field relaxation rates in $Pr_{1-x}Nd_xOs_4Sb_{12}$, $x = 0$, 0.05, 0.10, 0.15, 0.20, and 0.25, from fits of Eq. (7) to the data. Triangles: Gaussian relaxation rate $\sigma$. Circles: exponential damping rate $\lambda$. $H_T = 200$ Oe for $x = 0$; 100 Oe for $x > 0$.

## IV. CONCLUSIONS

We have carried out a study of Pr-rich $Pr_{1-x}Nd_xOs_4Sb_{12}$ alloys, $x = 0$, 0.05, 0.10, 0.15, 0.20, and 0.25, using the $\mu$SR technique. The spontaneous internal field found in the end compound $PrOs_4Sb_{12}$ is rapidly suppressed by Nd doping, and is not observed for $x \geqslant 0.05$. This is a surprising result since the $Nd^{3+}$ ions possess $4f$ magnetic moments, for which any static magnetism would break time reversal symmetry. Broken TRS is possible without an associated internal magnetic field if there are no spontaneous defect- or surface-induced supercurrents [42] and also no associated spin supercurrents [43]. It is unlikely, however, that doping with magnetic impurities would lead to either of these situations. Thus the observed suppression of the spontaneous field is strong evidence that TRS is restored for low Nd concentrations.

At higher Nd concentrations the dynamic relaxation rate increases with decreasing temperature as $T \to 0$, suggestive of critical slowing as a magnetic phase transition is ap-proached. There is, however, no sign of static magnetism, ordered or disordered, down to the lowest temperatures of measurement. This is surprising, since linear extrapolation of the ferromagnetic Curie temperature $T_{\mathrm{Curie}}$ (Fig. 1) to low Nd concentrations yields estimated transitions at experimentally-attained temperatures [e.g., $T_{\mathrm{Curie}}(\mathrm{est.}) = 0.16$ K for $x = 0.2$].

The superfluid density from TF-$\mu$SR measurements is remarkably independent of $x$ up to $\sim$0.2, i.e., for more than 20% suppression of $T_c$ (Fig. 1). For $x = 0$ $\rho_s(0) \approx 10^{22}$ cm$^{-3}$ [27], or roughly 8 carriers per cubic unit cell. This is somewhat high compared to band-theoretical Fermi volumes [44], but no more than order-of-magnitude agreement can be expected given the approximations in the $\mu$SR analysis.

We conclude that in Pr-rich $\mathrm{Pr}_{1-x}\mathrm{Nd}_x\mathrm{Os}_4\mathrm{Sb}_{12}$ the absence of static $\mathrm{Nd}^{3+}$ magnetism, the slow suppression of superfluid density, and the slow change of $T_c$ with Nd doping all suggest an anomalously weak coupling between $\mathrm{Nd}^{3+}$ spins and conduction-band states.

### ACKNOWLEDGMENTS

We are grateful to the staff of ISIS and TRIUMF for their invaluable help during the experiments at these facilities. This research was supported at CSU Fresno by U.S. NSF DMR-1506677, at UC San Diego by U.S. DOE DE-FG02-04ER46105 and NSF DMR-1810310, at CSU Los Angeles by NSF DMR-1523588, at Fudan U. by the National Key Research and Development Program of China (Grant No. 2017YFA0303104), at Hokkaido U. by JSPS KAKENHI 26400342, 15K05882, and 15K21732, and at UC Riverside by the UCR Academic Senate.

[1] E. D. Bauer, N. A. Frederick, P.-C. Ho, V. S. Zapf, and M. B. Maple, Superconductivity and heavy fermion behavior in $\mathrm{PrOs}_4\mathrm{Sb}_{12}$, Phys. Rev. B **65**, 100506(R) (2002).

[2] M. Maple, N. Frederick, P.-C. Ho, W. Yuhasz, and T. Yanagisawa, Unconventional superconductivity and heavy fermion behavior in $\mathrm{PrOs}_4\mathrm{Sb}_{12}$, J. Supercond. Novel Magn. **19**, 299 (2006).

[3] E. A. Goremychkin, R. Osborn, E. D. Bauer, N. A. Frederick, W. M. Yuhasz, F. M. Woodward, and J. W. Lynn, Crystal Field Potential of $\mathrm{PrOs}_4\mathrm{Sb}_{12}$: Consequences for Superconductivity, Phys. Rev. Lett. **93**, 157003 (2004).

[4] G. Seyfarth, J. P. Brison, M.-A. Méasson, J. Flouquet, K. Izawa, Y. Matsuda, H. Sugawara, and H. Sato, Multiband Superconductivity in the Heavy Fermion Compound PrOs$_4$Sb$_{12}$, Phys. Rev. Lett. **95**, 107004 (2005).

[5] K. Izawa, Y. Nakajima, J. Goryo, Y. Matsuda, S. Osaki, H. Sugawara, H. Sato, P. Thalmeier, and K. Maki, Multiple Superconducting Phases in New Heavy Fermion Superconductor PrOs$_4$Sb$_{12}$, Phys. Rev. Lett. **90**, 117001 (2003).

[6] E. E. M. Chia, M. B. Salamon, H. Sugawara, and H. Sato, Probing the Superconducting Gap Symmetry of PrOs$_4$Sb$_{12}$: A Penetration Depth Study, Phys. Rev. Lett. **91**, 247003 (2003).

[7] K. Grube, S. Drobnik, C. Pfleiderer, H. von Löhneysen, E. D. Bauer, and M. B. Maple, Specific heat and ac susceptibility studies of the superconducting phase diagram of PrOs$_4$Sb$_{12}$, Phys. Rev. B **73**, 104503 (2006).

[8] Y. Aoki, A. Tsuchiya, T. Kanayama, S. R. Saha, H. Sugawara, H. Sato, W. Higemoto, A. Koda, K. Ohishi, K. Nishiyama, and R. Kadono, Time-Reversal Symmetry-Breaking Superconductivity in Heavy-Fermion PrOs$_4$Sb$_{12}$ Detected by Muon-Spin Relaxation, Phys. Rev. Lett. **91**, 067003 (2003).

[9] P.-C. Ho, N. A. Frederick, V. S. Zapf, E. D. Bauer, T. D. Do, M. B. M. A. D. Christianson, and A. H. Lacerda, High-field ordered and superconducting phases of the heavy fermion compound PrOs$_4$Sb$_{12}$, Phys. Rev. B **67**, 180508(R) (2003).

[10] M. Kohgi, K. Iwasa, M. Nakajima, N. Metok, S. Araki, N. Bernhoeft, J.-M. Mignot, A. Gukasov, H. Sato, Y. Aoki, and H. Sugawara, Evidence for Magnetic-Field-Induced Quadrupolar Ordering in the Heavy-Fermion Superconductor PrOs$_4$Sb$_{12}$, J. Phys. Soc. Jpn. **72**, 1002 (2003).

[11] M. Yogi, H. Kotegawa, G.-q. Zheng, Y. Kitaoka, S. Ohsaki, H. Sugawara, and H. Sato, Novel Phase Transition Near the Quantum critical Point in the Filled Skutterudite Compound CeOs$_4$Sb$_{12}$: an Sb-NQR Study, J. Phys. Soc. Jpn. **74**, 1950 (2005).

[12] E. D. Bauer, A. Ślebarski, E. J. Freeman, C. Sirvent, and M. B. Maple, Kondo insulating behaviour in the filled skutterudite compound CeOs$_4$Sb$_{12}$, J. Phys. Condens. Matter **13**, 4495 (2001).

[13] C. Yang, Z. Zhou, H. Wang, J. Hu, K. Iwasa, H. Sugawara, and H. Sato, Evidence of spin density wave of CeOs$_4$Sb$_{12}$, Rare Metals **25**, 550 (2006).

[14] C. P. Yang, H. Wang, J. F. Hu, K. Iwasa, K. Kohgi, H. Sugawara, and H. Sato, Antiferro-

magnetic oedering of $CeOs_4Sb_{12}$ below 1K, J. Phys. Chem. C. **111**, 2391 (2007).

[15] P.-C. Ho, W. M. Yuhasz, N. P. Butch, N. A. Frederick, T. A. Sayles, J. R. Jeffries, M. B. Maple, J. B. Betts, A. H. Lacerda, P. Rogl, and G. Giester, Ferromagnetism and possible heavy fermion behavior in single crystals of $NdOs_4Sb_{12}$, Phys. Rev. B **72**, 094410 (2005).

[16] W. Y. Yuhasz, N. A. Frederick, P.-C. Ho, N. P. Butch, B. J. Taylor, T. A. Sayles, M. B. Maple, J. B. Betts, A. H. Lacerda, P. Rogl, and G. Giester, Heavy fermion behavior, crystalline electric field effects, and weak ferromagnetism in $SmOs_4Sb_{12}$, Phys. Rev. B **71**, 104402 (2005).

[17] P.-C. Ho, T. Yanagisawa, W. M. Yuhasz, A. A. Dooraghi, C. C. Robinson, N. P. Butch, R. E. Baumbach, and M. B. Maple, Superconductivity, magnetic order, and quadrupolar order in the filled skutterudite system $Pr_{1-x}Nd_xOs_4Sb_{12}$, Phys. Rev. B **83**, 024511 (2011).

[18] D. E. MacLaughlin, P.-C. Ho, L. Shu, O. O. Bernal, S. Zhao, A. A. Dooraghi, T. Yanagisawa, M. B. Maple, and R. H. Fukuda, Muon spin rotation and relaxation in $Pr_{1-x}Nd_xOs_4Sb_{12}$: Magnetic and superconducting ground states, Phys. Rev. B **89**, 144419 (2014).

[19] P.-C. Ho, D. E. MacLaughlin, L. Shu, O. O. Bernal, S. Zhao, A. A. Dooraghi, T. Yanagisawa, M. B. Maple, and R. H. Fukuda, Muon spin rotation and relaxation in $Pr_{1-x}Nd_xOs_4Sb_{12}$: Paramagnetic states, Phys. Rev. B **89**, 235111 (2014).

[20] L. Shu, W. Higemoto, Y. Aoki, A. D. Hillier, K. Ohishi, K. Ishida, R. Kadono, A. Koda, O. O. Bernal, D. E. MacLaughlin, Y. Tunashima, Y. Yonezawa, S. Sanada, D. Kikuchi, H. Sato, H. Sugawara, T. U. Ito, and M. B. Maple, Suppression of time-reversal symmetry breaking superconductivity in $Pr(Os_{1-x}Ru_x)_4Sb_{12}$ and $Pr_{1-y}La_yOs_4Sb_{12}$, Phys. Rev. B **83**, 100504 (2011).

[21] A. Schenck, *Muon Spin Rotation Spectroscopy: Principles and Applications in Solid State Physics* (A. Hilger, Bristol & Boston, 1985).

[22] J. H. Brewer, Muon Spin Rotation/Relaxation/Resonance, in *Encyclopedia of Applied Physics*, Vol. 11, edited by G. L. Trigg, E. S. Vera, and W. Greulich (VCH Publishers, New York, 1994) pp. 23–53; in *Digital Encyclopedia of Applied Physics*, edited by G. L. Trigg, E. S. Vera, and W. Greulich (Wiley-VCH Verlag GmbH & Co KGaA, Weinheim, 2003).

[23] S. J. Blundell, Spin-polarized muons in condensed matter physics, Contemp. Phys. **40**, 175 (1999).

[24] S. L. Lee, S. H. Kilcoyne, and R. Cywinski, eds., *Muon Science: Muons in Physics, Chemistry and Materials*, Scottish Universities Summer School in Physics No. 51 (Institute of Physics

Publishing, Bristol & Philadelphia, 1999).

[25] A. Yaouanc and P. Dalmas de Réotier, *Muon Spin Rotation, Relaxation, and Resonance: Applications to Condensed Matter*, International Series of Monographs on Physics (Oxford University Press, New York, 2011).

[26] W. Jeitschko and D. Braun, LaFe$_4$P$_{12}$ with filled CoAs$_3$-type structure and Isotypic lanthanoid-transition metal polyphosphides, Acta Cryst. B **33**, 3401 (1977).

[27] D. E. MacLaughlin, J. E. Sonier, R. H. Heffner, O. O. Bernal, B.-L. Young, M. S. Rose, G. D. Morris, E. D. Bauer, T. D. Do, and M. B. Maple, Muon Spin Relaxation and Isotropic Pairing in Superconducting PrOs$_4$Sb$_{12}$, Phys. Rev. Lett. **89**, 157001 (2002).

[28] L. Shu, D. E. MacLaughlin, Y. Aoki, Y. Tunashima, Y. Yonezawa, S. Sanada, D. Kikuchi, H. Sato, R. H. Heffner, W. Higemoto, K. Ohishi, T. U. Ito, O. O. Bernal, A. D. Hillier, R. Kadono, A. Koda, K. Ishida, H. Sugawara, N. A. Frederick, W. M. Yuhasz, T. A. Sayles, T. Yanagisawa, and M. B. Maple, Muon spin relaxation and hyperfine-enhanced $^{141}$Pr nuclear spin dynamics in Pr(Os,Ru)$_4$Sb$_{12}$ and (Pr,La)Os$_4$Sb$_{12}$, Phys. Rev. B **76**, 014527 (2007).

[29] L. Shu, D. E. MacLaughlin, W. P. Beyermann, R. H. Heffner, G. D. Morris, O. O. Bernal, F. D. Callaghan, J. E. Sonier, W. M. Yuhasz, N. A. Frederick, and M. B. Maple, Penetration depth, multiband superconductivity, and absence of muon-induced perturbation in superconducting PrOs$_4$Sb$_{12}$, Phys. Rev. B **79**, 174511 (2009).

[30] T. M. Riseman, J. H. Brewer, and D. J. Arseneau, Corrected asymmetry plots, Hyperfine Interact. **87**, 1135 (1994).

[31] R. Kubo and T. Toyabe, A stochastic model for low field resonance and relaxation, in *Magnetic Resonance and Relaxation*, edited by R. Blinc (North-Holland, Amsterdam, 1967) pp. 810–823; R. S. Hayano, Y. J. Uemura, J. Imazato, N. Nishida, T. Yamazaki, and R. Kubo, Zero- and low-field spin relaxation studied by positive muons, Phys. Rev. B **20**, 850 (1979).

[32] R. E. Walstedt and L. R. Walker, Nuclear-resonance line shapes due to magnetic impurities in metals, Phys. Rev. B **9**, 4857 (1974).

[33] Y. J. Uemura, T. Yamazaki, D. R. Harshman, M. Senba, and E. J. Ansaldo, Muon-spin relaxation in *Au*Fe and *Cu*Mn spin glasses, Phys. Rev. B **31**, 546 (1985).

[34] R. Kubo, A stochastic theory of spin relaxation, Hyperfine Interact. **8**, 731 (1981).

[35] M. R. Crook and R. Cywinski, Voigtian Kubo-Toyabe muon spin relaxation, J. Phys. Condens. Matter **9**, 1149 (1997).

[36] A. Maisuradze, W. Schnelle, R. Khasanov, R. Gumeniuk, M. Nicklas, H. Rosner, A. Leithe-Jasper, Y. Grin, A. Amato, and P. Thalmeier, Evidence for time-reversal symmetry breaking in superconducting PrPt$_4$Ge$_{12}$, Phys. Rev. B **82**, 024524 (2010).

[37] Y. Aoki, T. Tayama, T. Sakakibara, K. Kuwahara, K. Iwasa, M. Kohgi, W. Higemoto, D. E. MacLaughlin, H. Sugawara, and H. Sato, The Unconventional Superconductivity of Skutterudite PrOs$_4$Sb$_{12}$: Time-Reversal Symmetry Breaking and Adjacent Field-Induced Quadrupole Ordering, J. Phys. Soc. Jpn. **76**, 051006 (2007).

[38] J. E. Sonier, J. H. Brewer, and R. F. Kiefl, $\mu$SR studies of the vortex state in type-II superconductors, Rev. Mod. Phys. **72**, 769 (2000).

[39] J. E. Sonier, Muon spin rotation studies of electronic excitations and magnetism in the vortex cores of superconductors, Rep. Prog. Phys. **70**, 1717 (2007).

[40] E. H. Brandt, Flux distribution and penetration depth measured by muon spin rotation in high-$T_c$ superconductors, Phys. Rev. B **37**, 2349 (1988).

[41] L. C. Hebel and C. P. Slichter, Nuclear spin relaxation in normal and superconducting aluminum, Phys. Rev. **113**, 1504 (1959).

[42] See, e.g., M. Sigrist and K. Ueda, Phenomenological theory of unconventional superconductivity, Rev. Mod. Phys. **63**, 239 (1991), and references therein.

[43] T. Ohmi and K. Machida, Nonunitary Superconducting State in UPt$_3$, Phys. Rev. Lett. **71**, 625 (1993).

[44] H. Sugawara, S. Ōsaki, S. R. Saha, Y. Aoki, H. Sato, Y. Inada, H. Shishido, R. Settai, Y. Ōnuki, H. Harima, and K. Oikawa, Fermi surface of the heavy-fermion superconductor PrOs$_4$Sb$_{12}$, Phys. Rev. B **66**, 220504(R) (2002).